\documentclass[aps,showpacs,preprint,superscriptaddress]{revtex4}
\usepackage{graphicx}
\usepackage{amsmath}
\usepackage{ulem}
\usepackage{bm}
\begin{document}
\title{Pair production in strong SU(2) background fields}
\author{M. R. Jia}
\author{Z. L. Li}
\author{C. Lv}
\author{F. Wan}
\affiliation{College of Nuclear Science and Technology, Beijing Normal University, Beijing 100875, China}
\author{B. S. Xie \footnote{Corresponding author. Email address: bsxie@bnu.edu.cn}}
\affiliation{College of Nuclear Science and Technology, Beijing Normal University, Beijing 100875, China}
\affiliation{Beijing Radiation Center, Beijing 100875, China}

\date{\today}

\begin{abstract}
The fermion particle pair production in strong SU(2) gauged chromoelectric fields is studied by using Boltzmann-Vlasov equation in a classical way. The existence of pre-production process in a classical description is shown with the distribution evolution of non-Abelian particle production. It is interesting to find that the distribution center of particle number density is on two islands and has a split on color charge sphere as it evolutes and reaches a steady state at last, which is related to the amplitude and the varying of the field.

\end{abstract}
\maketitle

\section{Introduction}

The experiments of high energy collision of heavy ions are carrying out in both Relativistic Heavy Ion Collider (RHIC) and Large Hadron Collider (LHC), which are thought to produce a quark-gluon plasma. Efforts to model this plasma using a full quantum field theory have been proved particularly difficult. Since it is hard to deal with particles of different colors for different flavors under the gauge invariance and the renormalization, an appropriate approximation theory is needed urgently. Then a series of work has been carried out in effective classical theories \cite{Heinz1,Heinz2,Heinz3,Kelly1,Kelly2,Bodeker1,Bodeker2,Bodeker3}, and summarized in Ref.\cite{Daniel}.

For the problem of particle production, Schwinger developed a proper time method to calculate the particle production via tunneling mechanism in Quantum Electrodynamics (QED) theory \cite{Shwinger}. For QED case, Dawson and Cooper have studied particle production in semiclassical approach based on a Boltzmann-Vlasov (B-V) transport equation with an instantaneous source term \cite{Cooper20091,Cooper20092} and compared the B-V results with numerical simulations of the full quantum treatment. We have also made some researches on the relevant pair production problem in QED case. For instances, we examined carefully the effects of laser pulse envelop shape and carrier phases on produced pairs \cite{Xie1}. Also recently we studied the nonperturbative signatures on particle momentum spectra of pair production by solving the quantum Vlasov equation or in the Dirac-Heisenberg-Wigner formalism, especially for general elliptic polarized fields \cite{Xie2,Xie3,Xie4}. On the other hand, naturally, the problem of particle production is expected to be extended to the Quantum Chromodynamics (QCD) case. For example, the soft-gluon production \cite{Nayak and Nie} and the quark-antiquark production \cite{Nayak} in a constant chromoelectric background field were studied by Nayak and his coauthor. Recently, Dawson, Mihaila and Cooper have researched this problem in which the backreaction effects is included \cite{Dawson1}. They had also studied the Casimir dependence of transverse distribution in non-Abelian particle production \cite{Dawson2}. Similar work in the SU(2) case has also been carried out by Skokov and Le\'va by using the Wigner function formalism \cite{Skokov and Levai}.

Although it is important to study the final momentum distribution or the energy spectrum, it seems also necessary and meaningful to study the process of the non-Abelian particle production for a better understanding of particle production dynamics. For QED case there exist such studies \cite{Blaschke}. For QCD case, however, there is still lacking of corresponding studies. In this paper, therefore, we try to fill the gap while it is a little primary study. We will study the process of particle production in classical way by using the B-V equation. Meanwhile the Wong's equations in Ref.\cite{Wong} are used as motion equations of classical colored point particles. Since the non-Abelian particle production is related to the Casimir invariants, we will give the distribution of particle production on color charge $\mathbf{Q}$ sphere in SU(2) case. And we will discuss the evolution of the distribution in different background fields. Finally, we will give the total number of particle production.

For simplicity we treat the particles as point particles with classical color and the backreaction effect is ignored since its effect is an order lower than the background field \cite{Dawson1} under the assumption of low density approximation. Note that in the realistic situation the collision effect should be taken into account. However, in our work, a simplicity or/and lacking of enough reality of the model, in our classical calculation, the pre-production process we found, which is our main point, is still exhibiting some interesting features. And we think it is helpful to deep the understandings of relevant problem. On the other hand we will also use the Schwinger source when introducing different fields as in Ref.\cite{xue} since that the field can be regards as varying slowly enough.

The paper is organized as follows. In Sec.\uppercase\expandafter{\romannumeral2} we introduce the Wong's equation in semiclassical transport theory. In Sec.\uppercase\expandafter{\romannumeral3} we give the definition of distribution function with the measures in phase space. In Sec.\uppercase\expandafter{\romannumeral4} we derive the B-V equation and introduce a classical source term. In Sec.\uppercase\expandafter{\romannumeral5} we give the numerical methods and numerical results in a special case. And in the final section, we summarize our numerical results and give the conclusions. By the way in Appendix, we give some derivation of the Wong's equations and the solution of the B-V equation by using the method of characteristics.

\section{Wong's equations in semi-classical theory}

A theory describing the Yang-Mills (YM) field and isotopic-spin-carrying particles in classical limit has been derived by Wong in Ref.\cite{Wong}. Within microscopic description, the trajectories in space are known exactly, which read as
\begin{eqnarray}
&& m\frac{d\hat{x}^{\mu}}{dt}=\hat{p^\mu}\label{eq1},\\
&& m\frac{d\hat{p^\mu}}{dt}=g\hat{Q}^aF^{\mu\nu}_a(\hat{x})\hat{p}_\nu\label{eq2},\\
&& m\frac{d\hat{Q}^a}{dt}=gf^{abc}A^b_\mu(\hat{x})\hat{Q}^c\hat{p}^\mu\label{eq3},
\end{eqnarray}
where
\begin{equation}
\begin{aligned}
F^{a,\mu\nu}(x)=\partial^\mu A^{a,\nu}(x)&-\partial^\nu A^{a,\mu}(x)\\
&+gf^{abc}A^{b,\mu}(x)A^{c,\nu}(x)\label{eq4}
\end{aligned}
\end{equation}
is a general YM field and satisfies
\begin{equation}
D^{a,c}_\mu(x)F^{c,\mu\nu}(x)=\langle J^{a,\nu}(x)\rangle.\label{eq5}
\end{equation}
After some approximation and introducing the proper time $\tau$ for particles, the equations read
\begin{eqnarray}
&&m\frac{dx^\mu}{d\tau}=p^\mu\label{eq6},\\
&&m\frac{dp^\mu}{d\tau}=gQ^aF^{\mu\nu}_ap_\nu\label{eq7},\\
&&m\frac{dQ^a}{d\tau}=gf^{abc}A^b_\mu Q^c p^\mu\label{eq8}.
\end{eqnarray}
The equations are just made replacement by $ \hat{x^\mu}\rightarrow x^\mu, \hat{p^\mu}\rightarrow p^\mu, \hat{Q}^a\rightarrow Q^a$, and will be used as the leading order of motions for point particles in soft gauged fields.

\section{Microscopic distribution function}

To describe the ensemble of particles in phase space, it is convenient to define distribution function which depend on whole set of coordinates, $\mathbf{x},\mathbf{p},\mathbf{Q}$. Here, we introduce the one-particle distribution function $f(\mathbf{x},\mathbf{p},\mathbf{Q})$ and $n(\mathbf{x},\mathbf{p},\mathbf{Q})$, then the color current density energy-momentum tensor and particle number density are given by
\begin{eqnarray}
&& t^{\mu\nu}(\mathbf{x})=\int{d\mathbf{P}d\mathbf{Q}p^\mu p^\nu f(\mathbf{x},\mathbf{p},\mathbf{Q})}\label{eq9},\\
&& j^\mu_a(\mathbf{x})=\int{d\mathbf{P}d\mathbf{Q}p^\mu Q_a f(\mathbf{x},\mathbf{p},\mathbf{Q})}\label{eq10},\\
&& n(\mathbf{x})=\int{d\mathbf{P}d\mathbf{Q}f(\mathbf{x},\mathbf{p},\mathbf{Q})}\label{eq11},
\end{eqnarray}
where
\begin{eqnarray}
&& d\mathbf{P}=d^4p \theta(p_0)\delta(p^2-m^2),\label{eq12}\\
&& d\mathbf{Q}=d^3Q c_R \delta(Q_aQ_b-q_2),\label{eq13}
\end{eqnarray}
are momentum measure and SU(2) group measure which constraint the mass on-shell condition, positive energy and conservation of the group Casimirs. And $c_R$ and $q_2$ are parameters related to particles and their representations. Here, different from that in Ref.\cite{Heinz1}, we just define one distribution function for both particles and antiparticles in SU(2) gauge, which are distinguished by color charge $\mathbf{Q}$. From Wong's equations in SU(2) case we can write $\mathbf{Q}$ in spherical coordinates
\begin{eqnarray}
&& Q_1=J \sin\theta \cos\phi\label{eq14},\\
&& Q_2=J \sin\theta \sin\phi\label{eq15}\\
&& Q_3=J \cos\theta\label{eq16},
\end{eqnarray}
and we can find that $c_R=2/(\sqrt{3}\pi)$, $q_2=3/4$ and $J=\sqrt{3}/2$ in Refs.\cite{Dawson1,Daniel}.

\section{Bolzman-Vlasov equation with a classical source term}

\subsection{Boltzmann-Vlasov equation}

The B-V equation can be derived by considering the total derivative of the distribution function $f(\mathbf{x},\mathbf{p},\mathbf{Q})$. Considering the proper time $\tau$ we have
\begin{equation}
m\frac{df(\mathbf{x},\mathbf{p},\mathbf{Q})}{d\tau}=p^\mu\\ \mathcal{B}_\mu [\mathbf{A}]((\tau)f(\mathbf{x},\mathbf{p},\mathbf{Q})\label{eq17},
\end{equation}
where $\mathbf{B}_\mu [\mathbf{A}]((\tau)=\mathcal{D}_\mu [\mathbf{A}]-g\mathbf{Q}\cdot \mathbf{F}_{\mu\nu}(\tau)\partial_{p_\nu}$ is the the B-V differential operator with $\mathcal{D}_\mu[\mathbf{A}] = \partial_\mu - g \mathbf{A}(\tau)\cdot \mathbf{Q}\times \partial_\mu$  is a color-covariant derivative operator. So now requiring  that $p$ to be on-shell and $\mathbf{Q}$ satisfies the Casimir relation, the B-V equation is given by
\begin{equation}
p_\mu \mathbf{B}[\mathbf{A}](\tau)f(\mathbf{x},\mathbf{p},\mathbf{Q})=p^t \mathbf{C}(\mathbf{x},\mathbf{p},\mathbf{Q})\label{eq18}.
\end{equation}
Here we ignore the collision effects, and $C(\mathbf{x},\mathbf{p},\mathbf{Q})$ is a source term. Inserting Eqs.(\ref{eq6})-(\ref{eq8}) into Eq.(\ref{eq17}), the B-V equation becomes
\begin{equation}
\begin{aligned}
(p^\mu \frac{\partial}{\partial x^\mu}+g Q^a F^{\mu\nu}_a p_\nu \frac{\partial}{\partial p^\mu} & + f_{abc}A^b_\mu Q^c \frac{\partial}{\partial Q_a})f(\mathbf{x},\mathbf{p},\mathbf{Q})\\
		& =p^t \mathbf{C}(\mathbf{x},\mathbf{p},\mathbf{Q})\label{eq19}.
\end{aligned}
\end{equation}

\subsection{Particle production}

By using one-loop approximation of the quantum field theory, the particle production rate for fermions \cite{Nayak and Nie} is
\begin{equation}
\frac{d N_{q,\bar{q}}}{dt d^3x d^2 p_T}=- \frac{1}{4\pi^3}\sum\limits_{j=1}^{3}|g \lambda_j|\ln[1-exp(-\pi(p^2_T + m^2)/|g \lambda_j)]\label{eq20},
\end{equation}
where the $m$ is the mass of the quark and $\lambda_1$, $\lambda_2$, $\lambda_3$ are the gauge invariant eigenvalues. All these calculations are based on the Schwinger tunneling effect in a constant chromoelectric field.

For classical particle production we can make replacement by $E^a(t)T^a \rightarrow \mathbf{E}(t)\cdot \mathbf{Q}$ and replace the sum by integration $\mathbf{Q}$ as in Ref.\cite{Dawson1}. For a chromoelectric field which is varying slowly we can also get a $\mathbf{Q}$-dependent particle production rate $\mathbf{C}(\mathbf{x},\mathbf{p},\mathbf{Q})$ which is given by
\begin{equation}
\mathbf{C}(\mathbf{x},\mathbf{p},\mathbf{Q})=|\mathbf{Q} \cdot \mathbf{E}(\mathbf{x})|\mathbf{R}(\mathbf{x},\mathbf{p},\mathbf{Q})\delta(p_z)\label{eq21},
\end{equation}
with
\begin{equation}
\mathbf{R}(\mathbf{x},\mathbf{p},\mathbf{Q})=\mathbf{P}(\mathbf{x},\mathbf{p},\mathbf{Q}) \mathbf{S}(\mathbf{x},\mathbf{p},\mathbf{Q})\label{eq22},
\end{equation}
where
\begin{equation}
\mathbf{P}(\mathbf{x},\mathbf{p},\mathbf{Q})=1-2 f_0(\mathbf{x},\mathbf{p},\mathbf{Q})\label{eq23},
\end{equation}
and
\begin{equation}
\mathbf{S}(\mathbf{x},\mathbf{p},\mathbf{Q})=[1-exp(-[\frac{\pi(\mathbf{p}^2 +m^2)}{|g\mathbf{Q}\cdot \mathbf{E}(\mathbf{x})|}])]\label{eq24}.
\end{equation}
Here the $f_0(\mathbf{x},\mathbf{p},\mathbf{Q})$ is the special distribution function which is defined as $f_0(\mathbf{x},\mathbf{p},\mathbf{Q})\equiv f(\mathbf{x},\mathbf{p}(p_z=0),\mathbf{Q})$. It is noted that most of the particle production occurs at $p_z=0$ by the Pauli-Blocking at $p_z$.

\section{Numerical results}

In this section we give the numerical methods and results. To simplify the calculation, we choose the gauged potential $A^{a,\mu}(x)$ in $z$ direction depending only on $t$ for all $a =1,2,3$ which read as
\begin{equation}
A^{a,\mu}(t)=(0,0,0,A^a(t))\label{eq25}.
\end{equation}
To calculate the distribution of particle production on the $\mathbf{Q}$ sphere, we should solve the trajectory equations numerically for the distribution function. Since no particles produced at $t=0$, we set $n(0,0,Q)=0$. We ignore the Pauli-Blocking effects in our calculation and set $g=1$ and use the unit of $m=1$. We only consider the particles created with zero momentum of $z$, then the initial conditions are given as
\begin{eqnarray}
t(\tau_0)=0 &, p^t(\tau_0)&=m\label{eq26},\\
z(\tau_0)=0 &, p^z(\tau_0)&=0\label{eq27}.
\end{eqnarray}
For the non-Abelian gauged vector potential we set $A=(0,0,0)$ at the beginning. We calculate both constant and cosine chromoelectric fields after giving their amplitude $E_{\mathbf{const}}$ and $E_{\mathbf{cos}}$, and we appropriately choose the constant fields equal their amplitude and the frequency of cosine fields is $0.8$ in unit of $m=1$.

\begin{figure}
\centering
\includegraphics[width=15cm]{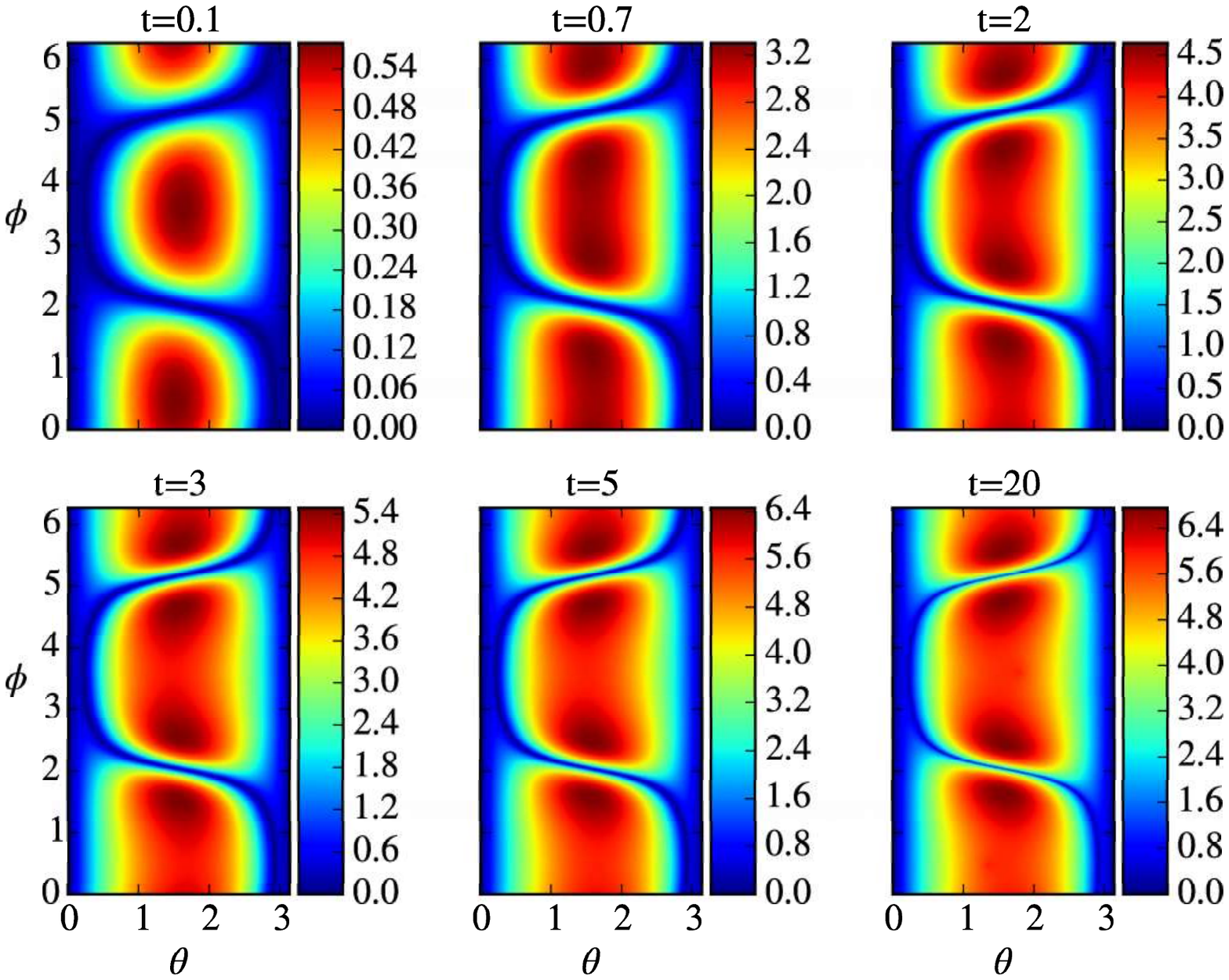}
%\hspace{1in}
\caption{(color online). Contour of produced particle number density on $\mathbf{Q}$ sphere in a constant chromoelectric field at different time. The color bar has been multiplied by $10^{-5}$.}
\label{fig1}
\end{figure}

\begin{figure}
\centering
\includegraphics[width=15cm]{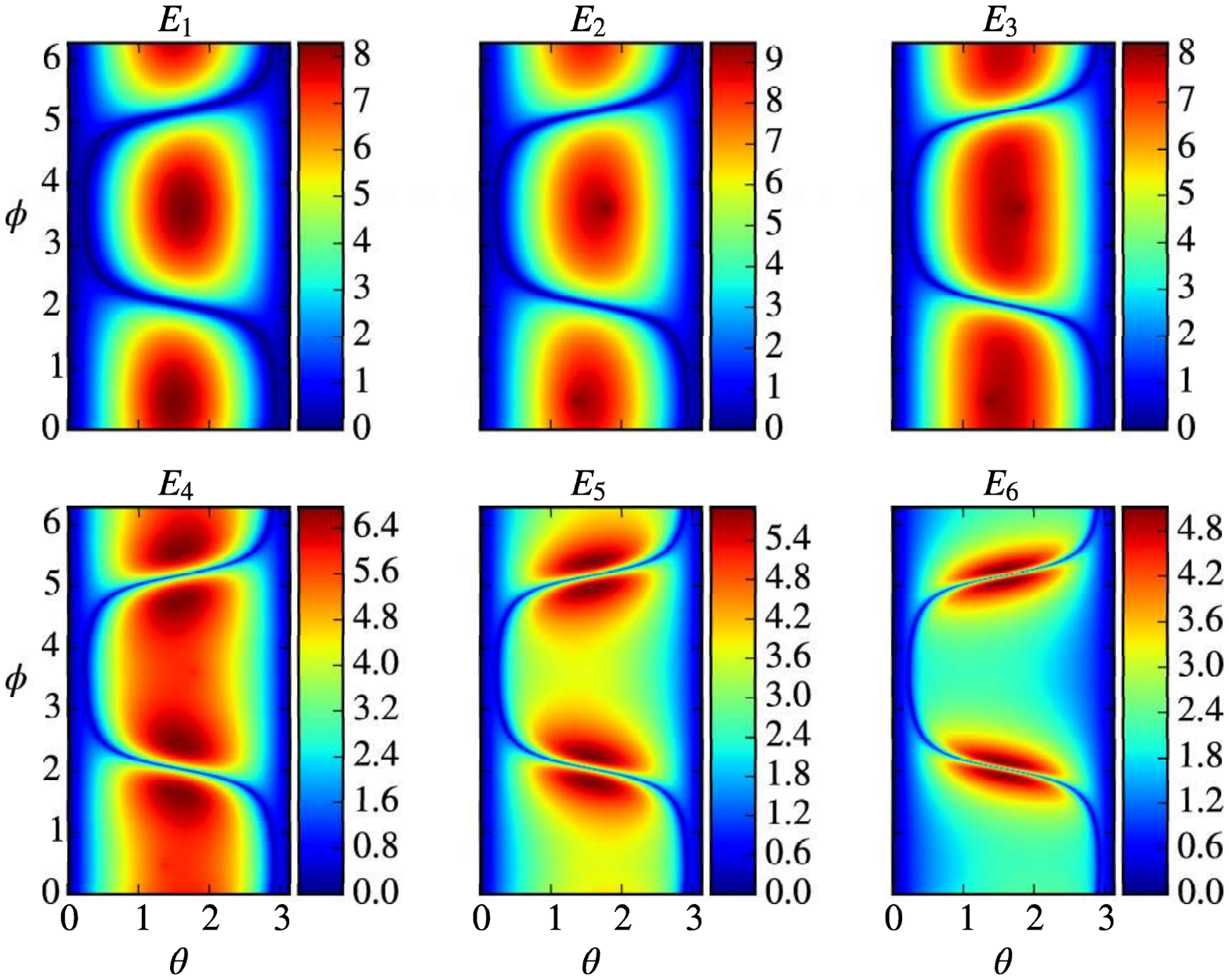}
%\hspace{1in}
\caption{(color online). Contour of produced particle number density on $\mathbf{Q}$ sphere in constant chromoelectric fields at $t=20$ for different field amplitude of $E_1=(1,0.5,0.25)$, $E_2=(2,1,0.5)$, $E_3=(3,1.5,0.75)$, $E_4=(4,2,1)$, $E_5=(8,4,2)$, and $E_6=(16,8,4)$, respectively. The color bar has been multiplied by $10^{-5}$.}
\label{fig2}
\end{figure}

\begin{figure}
\centering
\includegraphics[width=15cm]{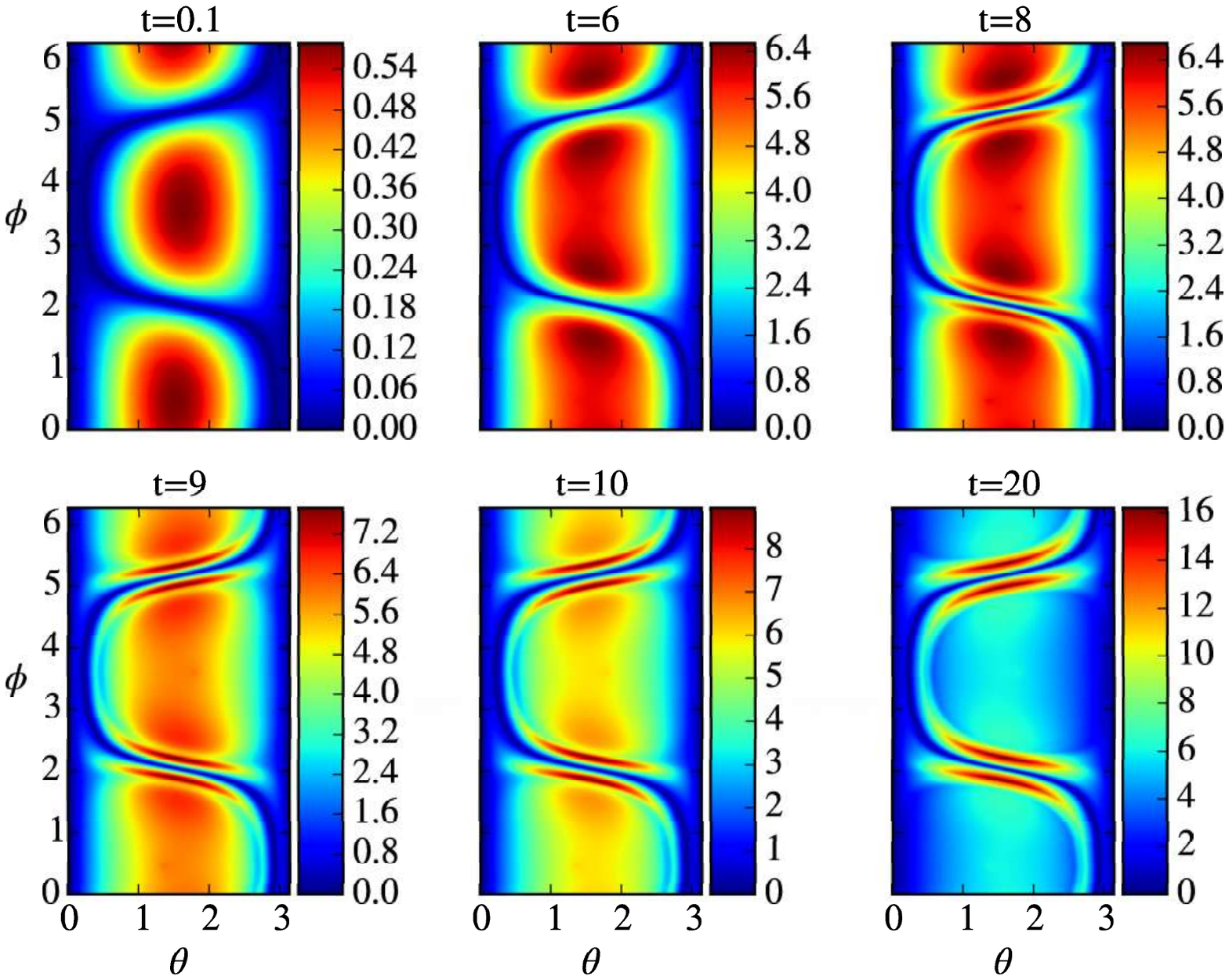}
%\hspace{1in}
\caption{(color online). The same as in Fig.\ref{fig1} except in a cosine chromoelectric field.}
\label{fig3}
\end{figure}

\begin{figure}
\centering
\includegraphics[width=15cm]{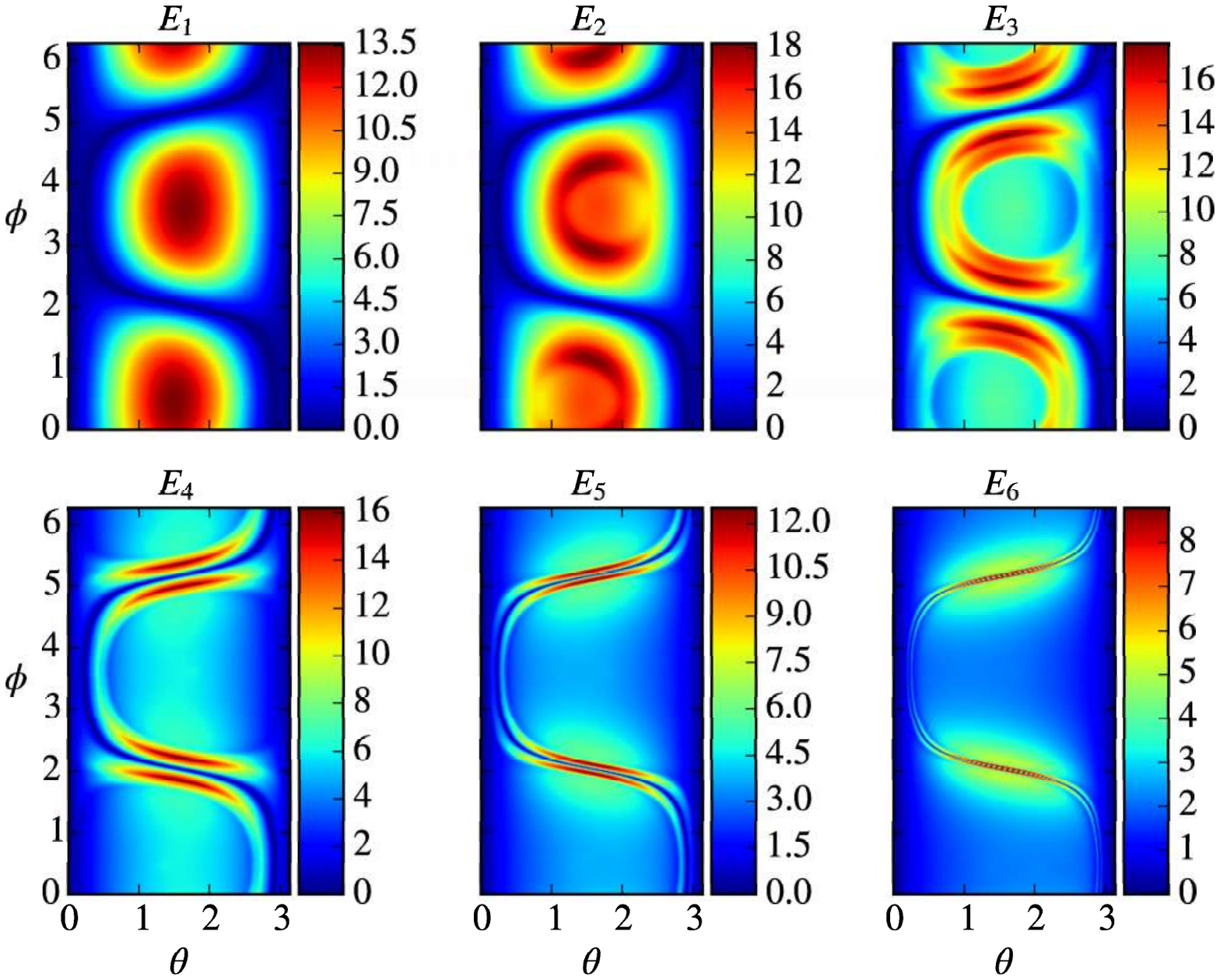}
%\hspace{1in}
\caption{(color online). The same as in Fig.\ref{fig2} except in a cosine chromoelectric field.}
\label{fig4}
\end{figure}

\begin{figure}
\centering
\includegraphics[width=12cm]{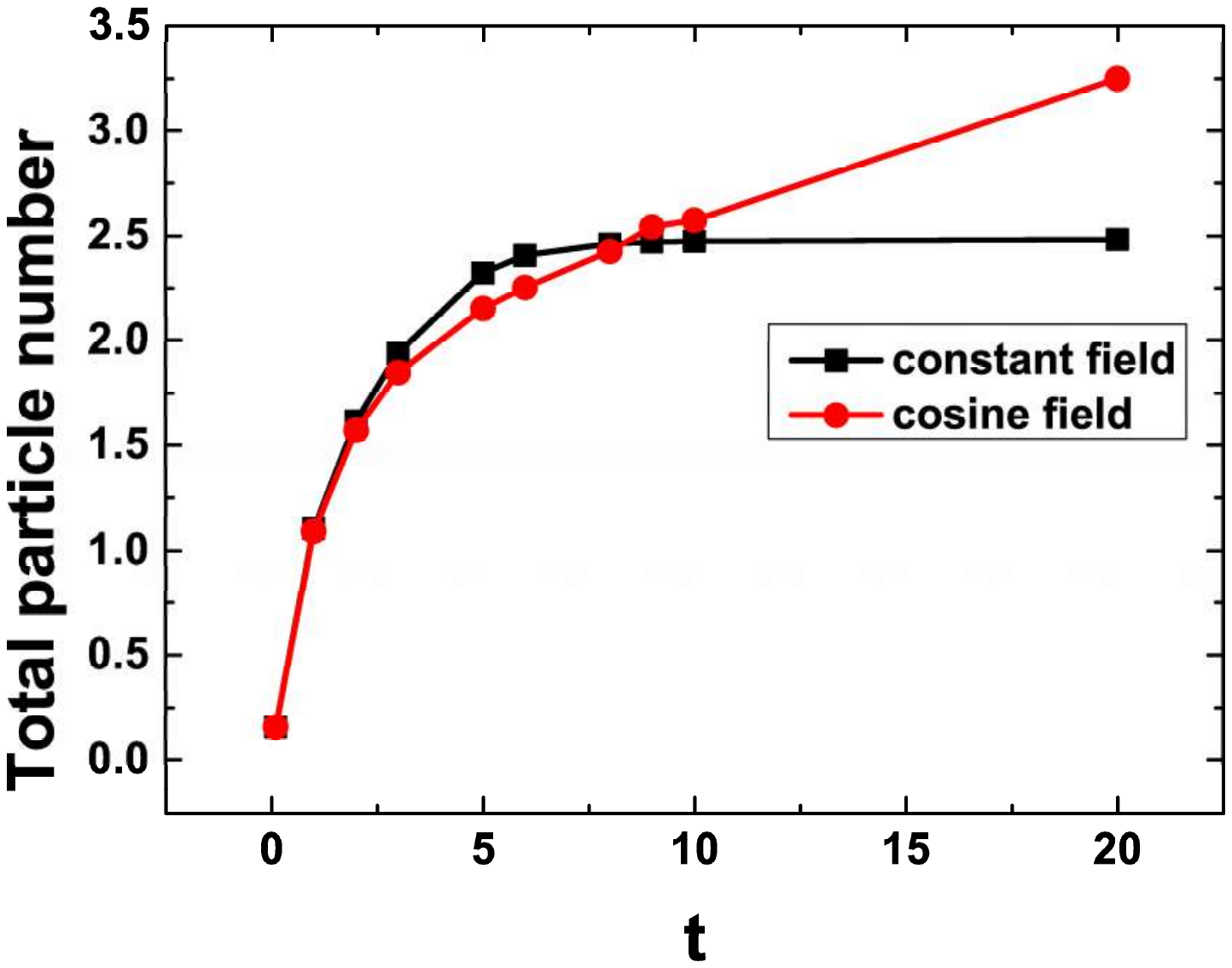}
%\hspace{1in}
\caption{(color online). Total produced particle number on $\mathbf{Q}$ sphere for fermions as a function of $t$.}
\label{fig5}
\end{figure}

\begin{figure}
\centering
\includegraphics[width=12cm]{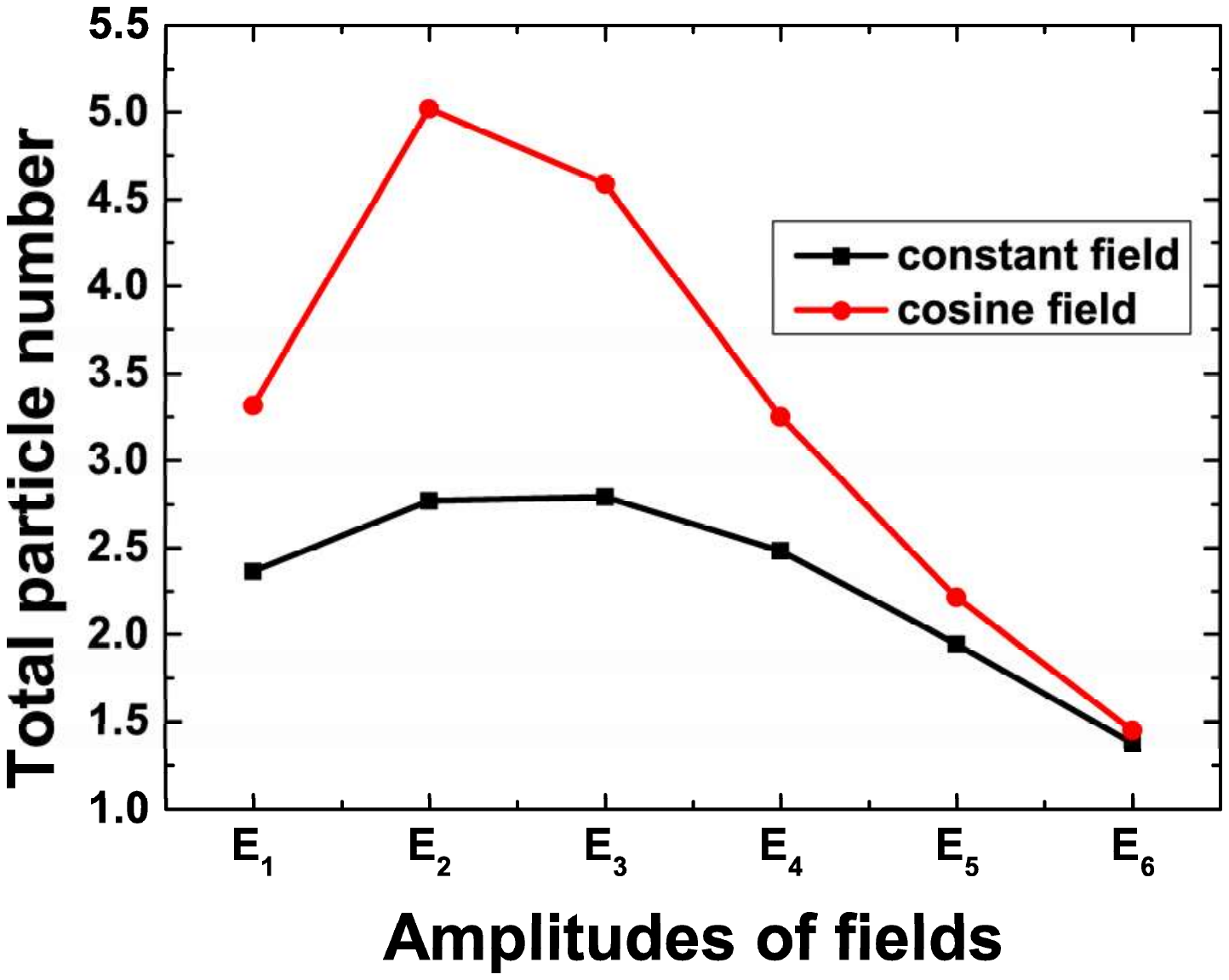}
%\hspace{1in}
\caption{(color online). Total produced particle number on $\mathbf{Q}$ sphere for fermions as a function of fields amplitude at $t=20$. The field amplitudes $E_i$ with $i=1$ to $i=6$ are the same as in Fig.\ref{fig2}.}
\label{fig6}
\end{figure}

We distribute the color charges on a $\mathbf{Q}$ sphere, and comprehend it as the non-Abelian interaction acts when $\mathbf{Q}$ moves on the sphere with the Wong's equations. We take each degree of $\theta$ and $\phi$ on $\mathbf{Q}$ sphere and calculate the produced particle number density as time goes on. We take $d\tau =0.0005$, and choose $\tau_0=0$ to guarantee the production occurs in the light cone. Also we fix the $\mathbf{Q}$ at each step to guarantee the Wong's equations won't be violated.

After solving the trajectory equations numerically, finding the distribution function, we can integrate out the produced particle number density $n(\mathbf{x},\mathbf{p},\mathbf{Q})$. And especially for SU(2) case we can distribute color charges on the $\mathbf{Q}$ sphere, and use $\theta$ and $\phi$ as two parameters to show the distribution of produced particle number density in contour maps.

In our calculation, we find that the number density distributes into two islands on $\mathbf{Q}$ sphere. Which we regard it as a pair of conjugate particles. And the distribution on the islands is center-symmetrical. Then we will concentrate on describing the phenomena on one island.

In Fig.\ref{fig1}, the distribution of number density is centered on the island at the beginning. And distribution center splits into two parts which go to each edge of the island as the time goes on. The distribution becomes steady when the time is long enough. In Fig.\ref{fig2}, we find that the steady state, the distribution could reach, is related to the amplitude of the fields. In small amplitude field the center won't split at all, but the split occurs when the amplitude become larger. And the two parts of the center distribute more closely to the edge as the amplitude of the field increases.

In Fig.\ref{fig3}, we can also find the same phenomenon in a slow varying field. The difference is that fluctuation structure appears in the contour map as the time goes on. And compared to the constant field, the split two parts are closer to the edge when they reach the steady state. Fig.\ref{fig4} shows the contour maps in cosine fields of different amplitude, it gives the same results from the constant field except the delicate fluctuation structure exists.

Our calculation on time is lasted to $t=70$, but we just show the contour maps which describe the splitting process of distribution center. In our calculation on time, we also find that the two parts of split distribution center are close to the edge, but never escape from the island. The calculation in fields of different amplitude also shows the confinement character. From Fig.\ref{fig2} and Fig.\ref{fig4}, it seems that there is a critical amplitude which leads to the distribution center splitting. Since it is hard to define the splitting strictly in classical calculation, we just choose some discrete amplitudes.

In Fig.\ref{fig5}, we show the total particle number evolution. We find that the total particle number increases quicker in constant field than in cosine field when $t<8$, but slower when $t>8$. Compare Fig.\ref{fig5} with Fig.\ref{fig1} and Fig.\ref{fig3}, we find that the distribution reach the steady state quicker but the total particle number in constant field is less than that in cosine field.

In Fig.\ref{fig6}, we show the total particle number in fields of different amplitude at $t=20$, which is long enough to form a steady distribution on $\mathbf{Q}$ sphere. We find that the total particle number increases at first but then decreases as the amplitude grows in both of constant and cosine field. And we find the most total particle number is in $E_2$ field, the least total particle number is in $E_6$ field,  but the distribution center splits inconspicuously in $E_2$ field, conspicuously in $E_6$ field. We can also find the total particle number in constant fields is lesser than that in cosine fields when the same amplitudes are given.

\section{Conclusion and discussion}

By calculating the distribution of particle production on $\mathbf{Q}$ sphere, it is interesting to find that the distribution of particle number density is on two islands and has a movement on $\mathbf{Q}$ sphere as it evolutes. The distribution is centered at the beginning, splits into two parts which go to each edge of the islands, and reaches a steady state at last. We comprehend it as a pre-production process in classical describing, which means we could comprehend the non-Abelian production dose not occur immediately but there is a process. And we regard the two center-symmetrical islands as the color charge space of conjugated particles. From our calculation, we also find that the distribution can reach, is related to the amplitude and the varying of fields in space-time.

The other important aspect is the Pauli-Blocking effects which is not included in present work. In fact we have made some calculation for the problem by including the Pauli-Blocking effects and it is found that the similar pre-production process exist. Certainly the difference of results quantitatively between two cases of with and without Pauli-Blocking effect is worthy to study further.

For future work, it seems necessary to define the splitting in a proper way, and find the value of the critical amplitude which leads to the splitting. Meanwhile, extending the problem to SU(3) case and making the comparison with quantum field theoretic results are also important and requirement.

\begin{acknowledgments}
This work was supported by the National Natural Science Foundation of China (NSFC) under Grant No. 11475026. The computation was carried out at the HSCC of the Beijing Normal University.
\end{acknowledgments}

\appendix

\section{motion equations}

To simplify the formalism, we choose a gauge and require the field $A^{a,\mu}(x)$ in $z$ direction depending only on $t$ for all $a=1,2,3$, which read as
\begin{equation}
A^{a,\mu}(t)=(0,0,0,A^a(t))\label{eqA1}.
\end{equation}
We can easily find the non-vanishing field terms when $\mu =z,\nu =t$ or $\mu =t,\nu =z$ which give
\begin{eqnarray}
 F^a_{t,z}(t) = -F^a_{z,t}(t) = -\partial_t A^a(t) \equiv E^a(t)\label{eqA2}.
\end{eqnarray}
From Eq.(\ref{eq5}), the non-vanishing terms are
\begin{eqnarray}
&& \partial_t F^{a,tz}(t)=-\partial E^a(t)\equiv J^{a,z}(t)\label{eqA3},\\
&& gf^{abc}A^b_z{t}F^{a,zt}(t)=gf^{abc}A^b_z{t}E^c(t) \equiv J^{a,t}(t)\label{eqA4}.
\end{eqnarray}
Writing these equations in vector notation
\begin{eqnarray}
\partial_t \mathbf{A}(t)=-\mathbf{E}(t)\label{eqA5},\\
\partial_t \mathbf{E}(t)=-\mathbf{J}^z(t)\label{eqA6},\\
g\mathbf{A}(t)\times \mathbf{E}(t)=\mathbf{J}^t(t)\label{eqA7},
\end{eqnarray}
and from these Eqs.(\ref{eqA5})-(\ref{eqA7}) we have
\begin{equation}
\partial_t \mathbf{J}^t(t)+ g\mathbf{A}(t)\times \mathbf{J}^z(t)=0\label{eqA8}.
\end{equation}
Inserting these equations into Eqs.(\ref{eq6})-(\ref{eq8}), the Wong's equation can be obtained as
\begin{eqnarray}
&& m\frac{dt(\tau)}{d\tau}=p^t(\tau)\label{eqA9},\\
&& m\frac{dz(\tau)}{d\tau}=p^z(\tau)\label{eqA10},\\
&& m\frac{dp^t(\tau)}{d\tau}=g\mathbf{Q}(\tau)\cdot \mathbf{E}(\tau)p^z(\tau)\label{eqA11},\\
&& m\frac{dp^z(\tau)}{d\tau}=g\mathbf{Q}(\tau)\cdot \mathbf{E}(\tau)p^t(\tau)\label{eqA12},\\
&& m\frac{d\mathbf{Q}(\tau)}{d\tau}=g\mathbf{A}(\tau)\times \mathbf{Q}(\tau) p^z(\tau)\label{eqA13}.
\end{eqnarray}
By a dot product with $\mathbf{Q}(\tau)$ in Eq.(\ref{eqA13}) we could easily get $Q^2$ is a conserved quantity. And by a dot product with $\mathbf{A}(\tau)$ in Eq.(\ref{eqA13}) we have
\begin{equation}
m\mathbf{A}(\tau)\cdot \frac{d\mathbf{Q}(\tau)}{d\tau}=0\label{eqA14}.
\end{equation}
Reminding that
\begin{eqnarray}
p^t(\tau)\mathbf{E}(\tau)=-m\frac{dt}{d\tau}\frac{\mathbf{A}(t)}{dt}=-m\frac{d\mathbf{A}(\tau)}{d\tau}\label{eqA15},
\end{eqnarray}
then Eq.(\ref{eqA12}) can be written as a total derivative
\begin{equation}
m\frac{d}{d\tau}[p^z(\tau)+g\mathbf{Q}(\tau)\cdot \mathbf{A}(\tau)]=0\label{eqA16},
\end{equation}
so that $p^z(\tau)+g\mathbf{Q}(\tau)\cdot \mathbf{A}(\tau)\equiv \mathbf{P}^z$ is a constant of motion.

\section{solution of the B-V equation}

We could solve the B-V equation by using the method of characteristics. Integrate the source over a classical particle path trajectory from $\tau_0$ to $\tau$, and require the field in $z$ direction and take $p^z$ distribution only, then the distribution function $f(t,p^z,\mathbf{Q})$ reads as
\begin{equation}
\begin{aligned}
f(t,p^z,\mathbf{Q})= & f(t(\tau_0),p^z(\tau_0),\mathbf{Q}(\tau_0))))\\
            & +\frac{1}{m}\int\limits_{\tau_0}^{\tau}d\tau' p^t(\tau')\mathbf{C}(t(\tau'),p^z(\tau'),\mathbf{Q}(\tau'))\label{eqB1},
\end{aligned}
\end{equation}
here $t(\tau'),p^z(\tau'),\mathbf{Q}(\tau')$ are the solutions of the trajectory equations for values between $\tau_0$ and $\tau$ where the path length is real and $\tau > 0$, which means the integration is in the light cone as the time goes on. Since no particles present at $t=0$ so $f_0(0,p^z(\tau_0),\mathbf{Q}(\tau_0)))=0$ and Eq.(\ref{eqB1}) becomes
\begin{equation}
f(t,p^z,\mathbf{Q})=\frac{1}{m}\int\limits_{\tau_0}^{\tau}d\tau' p^t(\tau')\mathbf{C}(t(\tau'),p^z(\tau'),\mathbf{Q}(\tau'))\label{eqB2}.
\end{equation}
Inserting the classical source term, we have
\begin{equation}
\begin{aligned}
f(t,p^z,\mathbf{Q})=\frac{1}{m}\int\limits_{\tau_0}^{\tau}d\tau' & p^t(\tau')|g\mathbf{Q}(\tau')\cdot E(\tau')| \\
& \times \mathbf{R}(\tau',\mathbf{Q}(\tau')) \delta(p^z (\tau'))\label{eqB3},
\end{aligned}
\end{equation}
considering Eqs.(\ref{eqA12}),(\ref{eqB1}) and integrating over $p^z(\tau')$ find
\begin{equation}
f(t(\tau),\mathbf{Q}(\tau))=\sum \mathbf{R}(\tau_n,\mathbf{Q}(\tau_n)))\theta(t(\tau_n))\theta(t-t(\tau_n))\label{eqB4},
\end{equation}
here $\tau_n$ are solutions of
\begin{equation}
\mathbf{P}^z+[\mathbf{Q}\cdot \mathbf{A}-\mathbf{Q}(\tau')\cdot \mathbf{A}(\tau')]=0\label{eqB5},
\end{equation}
using the relation $\theta(0)=\frac{1}{2}$, and solving Eq.(\ref{eqB4}) and defining $f_0(t,0,\mathbf{Q})\equiv f(t,p^z,\mathbf{Q})$ gives
\begin{equation}
f_0(t,\mathbf{Q})=\frac{\mathbf{S}(t,\mathbf{Q})/2+\sum_{\tau_n'}^{}\mathbf{R}(\tau_n',\mathbf{Q}(\tau)n'))}{1+\mathbf{S}(t,Q)}\label{eqB6},
\end{equation}
which completes the solution of $f(t,p,\mathbf{Q})$ using the method of characteristics. In the special case, once the trajectory equations are solved the special distribution function could be easily calculated. Substituting the results into Eqs.(\ref{eq9})-(\ref{eq11}) then the need current energy, pressures and number density are got
\begin{eqnarray}
&& J^z(t)=\int d\mathbf{P} d\mathbf{Q} p^z Q f_0(t,\mathbf{Q})\label{eqB7},\\
&& J^t(t)=\int d\mathbf{P} d\mathbf{Q} p^t Q f_0(t,\mathbf{Q})\label{eqB8},\\
&& e(t)=\int d\mathbf{P} d\mathbf{Q} {p^t}^2 f_0(t,\mathbf{Q})\label{eqB9},\\
&& p^z(t)=\int d\mathbf{P} d\mathbf{Q} p^t p^z f_0(t,\mathbf{Q})\label{eqB10},\\
&& n(t)=\int d\mathbf{P} d\mathbf{Q} f_0(t,\mathbf{Q})\label{eqB11}.
\end{eqnarray}

\end{document}